\documentclass[preprint,5p,times,twocolumn]{elsarticle}

\usepackage{xurl}
\usepackage[colorlinks]{hyperref}
\usepackage[USenglish]{babel}

\usepackage[dvipsnames]{xcolor}

\journal{NIM A}

\begin{document}

\begin{frontmatter}

\newcommand{\thetitle}{Developing a Network Discovery Protocol for the Constellation Control and Data Acquisition Framework}
\title{\thetitle}
\hypersetup{pdftitle={\thetitle}, pdfauthor={Stephan Lachnit}}

\author{Stephan Lachnit\corref{sl}}
\ead{stephan.lachnit@desy.de}
\author{on behalf of the EDDA collaboration}

\affiliation{
  organization={Deutsches Elektronen-Synchrotron DESY},
  addressline={Notkestr. 85},
  city={22607 Hamburg},
  country={Germany}
}

\cortext[sl]{Corresponding author}

\begin{abstract}

Qualifying new detectors in test beam environments presents a challenging setting that requires stable operation of diverse devices, often employing multiple data acquisition systems. Changes to these setups are frequent, such as using different reference detectors depending on the facility. Managing this complexity necessitates a system capable of controlling the data taking, monitoring the experimental setup, facilitating seamless configuration, and easy integration of new devices.

One aspect of such systems is network configuration. Many systems require fixed IP addresses for all machines participating in the data acquisition, which adds complexity for users.

In this paper, a network protocol for network discovery tailored towards network-distributed control and data acquisition systems is described.

\end{abstract}


\end{frontmatter}


\section{Introduction}

The characterization and testing of detectors during their development is essential for the success of particle physics experiments. One environment for such tests is test beams, where a particle detector is tested using particles provided by a particle accelerator. To study the performance of the detector, reference measurements of the particles are needed, which is typically achieved using other detectors provided by the test beam facility. This results in a dynamic environment where the components for the measurement frequently change.

Each detector requires a readout system to read the data from the detector. These readout systems are highly specific to the detector itself and often come with specialized Data Acquisition~(DAQ) software, forming a DAQ system. In dynamic environments different DAQ systems for the respective detectors have to be operated synchronously to take data. This requires a control system, which provides a common interface to the different DAQ systems.

Since the DAQ systems for these detectors usually run on separate machines, such a control system has to incorporate network communication. Existing control systems like \mbox{EUDAQ2}~\cite{Liu2019:EUDAQ2} or \mbox{DAQling}~\cite{Boretto2020:DAQling} require fixed IP addresses for the setup. As a result, the first configuration step during test beam after the physical setup is often configuring IP addresses and copying them to a script or configuration file.

This paper describes an alternative approach using zero-configuration networking to discover components of the control system in a local network automatically.

\section{Zero-Configuration Networking}

Requiring IP addresses to be fixed is not necessary in local networks. Zero-configuration networking (often called zeroconf, network discovery, or service discovery) describes technologies that enable an automatic network setup without prior knowledge of other IP addresses. It has existed in some form at least since the mid-1980s with AppleTalk~\cite{AppleTalk}.

In the 2000s, several zero-configuration networking protocols emerged such as Apple's Bonjour~\cite{Bonjour} or Universal Plug and Play (UPnP)~\cite{UPnP} based on the User Datagram Protocol (UDP)~\cite{UDP}.

Data over UDP can be transmitted via unicasts, multicasts, or broadcasts. Unicasts transport data from one peer to another. Multicasts transport data from one peer to all peers that have explicitly joined the multicast group to which the data was sent. This multicast group is a special IP address that has to be defined by network protocols, similar to ports. Broadcasts transport data to all peers in a local network. Both UDP multicasts and broadcasts are thus suitable for zero-configuration networking.

In 2013, DNS-based Service Discovery (DNS-SD)~\cite{DNS-SD} based on Bonjour was standardized and is now the dominant network discovery protocol. It is the underlying technology behind most network printers via Bonjour, Spotify Connect~\cite{SpotifyConnectZeroConfAPI}, the Matter IoT protocol~\cite{MatterDiscovery}, and many other applications. It is mostly used in conjunction with multicast DNS (mDNS)~\cite{mDNS}, which uses UDP multicasts. DNS-SD is natively supported in Linux~(avahi), MacOS~(Bonjour), and Windows~(part of \texttt{dnsapi} since Windows 10).

Many programs with a limited scope implement custom discovery protocols based on UDP since the implementation can be faster than using the different native interfaces which differ substantially. Embedded solutions are not suitable since running multiple mDNS responders on a machine can lead to issues since queries might be answered multiple times with inconsistent responses. 

\section{Constellation}

Constellation~\cite{ConstellationWebsite} is a network-distributed control and data acquisition framework for small-scale experiments and experimental setups with volatile and dynamic constituents such as test beam environments or laboratory test stands. The central components of a Constellation network are so-called satellites, which implement instrument-controlling code or other components that should follow the Constellation operation synchronously. Satellites operate autonomously, meaning no central control instance is required to be active at all times.

Since Constellation aims to provide a flexible framework that is easy to use for operators and due to the autonomy of the satellites, a network discovery protocol that avoids requiring fixed IP addresses is necessary. This protocol and its implementation will be described in the following section.

\section{The Protocol}

\subsection{Requirements}

The main task of the protocol is to provide a way to discover satellites with a Constellation. This has to work both as an early joiner (discover satellites joining later) and as a late joiner (discover already existing satellites).

In a lab, multiple independent experiments might take place on the same local network. Thus, the protocol needs to provide an identifier to separate between experiments such that only satellites of one particular experiment can be discovered. A unique identifier for instances and an identifier for the type of service is also required.

Since multiple satellites might run on the same machine, network ports for specific services cannot be predefined, since only one application can bind a port.\footnote{This is not true for UDP but does apply to TCP, which is used for all network communication in Constellation besides the network discovery.} Services have to use an ephemeral port, which is a free port assigned by the operation system. The network discovery protocol thus also needs to provide the port of the service.

\subsection{Specifications}

The protocol is called \emph{Constellation Host Identification and Reconnaissance Protocol} (CHIRP) and is written as an IETF RFC-style document~\cite{CHIRP}.

\begin{figure}
  \centering
  \includegraphics[width=0.8\linewidth]{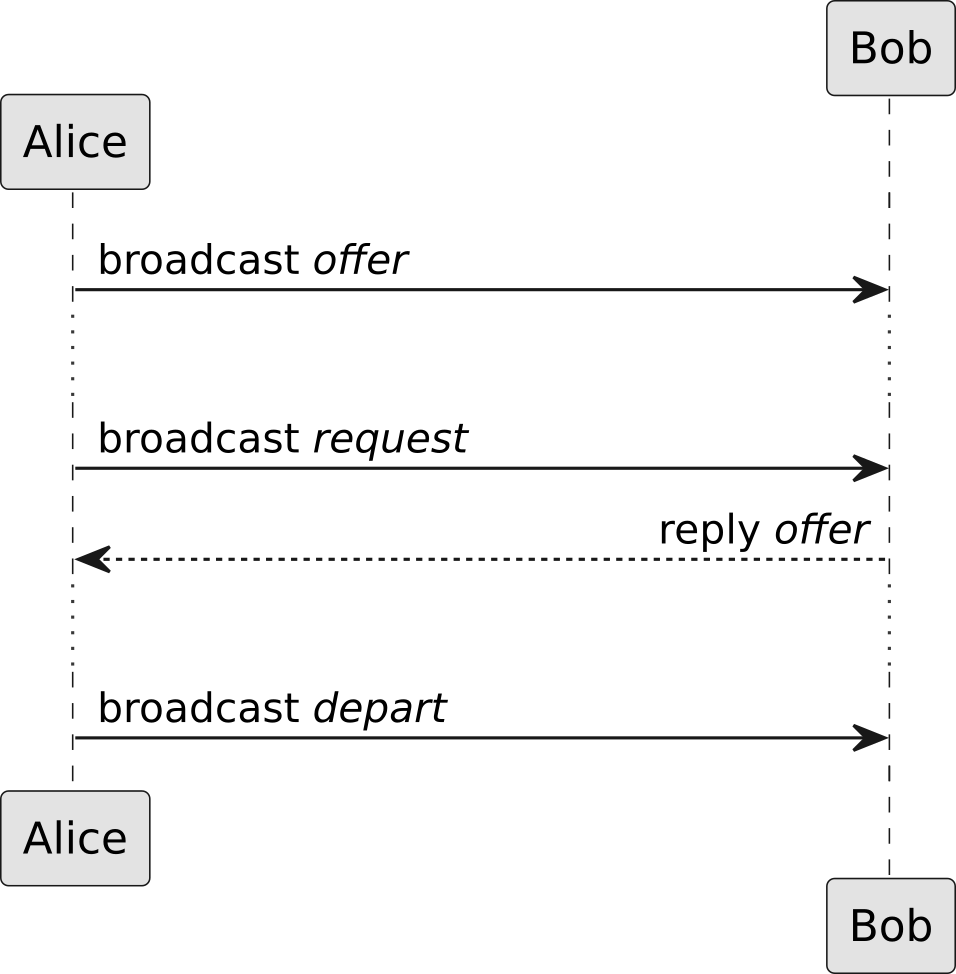}
  \caption{Sequence diagram for CHIRP}
  \label{fig:UmlCHIRP}
\end{figure}

CHIRP uses UDP broadcasts on port 7123. When a satellite is started, it broadcasts an \emph{offer} for each service it provides with the corresponding port number. To discover existing satellites, a \emph{request} can be broadcast, to which satellites reply with an \emph{offer}. Further, satellites should broadcast a \emph{depart} message when shutting down. \autoref{fig:UmlCHIRP} shows a sequence diagram of this logic.

Each host participating in CHIRP requires a 16-octet universally unique identifier (UUID). Each host also belongs to a group, which is identified by a 16-octet UUID. A host should only react to CHIRP messages from its corresponding group. The bit width of the host and group UUIDs was chosen to allow using MD5 hashes of arbitrary length names.

A CHIRP message has a fixed size of 42 octets. The first six octets are the protocol header, consisting of the five ASCII letters for the protocol~(\texttt{CHIRP}) and one octet for the protocol version~(\texttt{0x01}):

{\small
\begin{verbatim}
  +---+---+---+---+---+------+
  | C | H | I | R | P | 0x01 |
  +---+---+---+---+---+------+
\end{verbatim}
}

The body of a CHIRP message consists of a 1-octet message type, a 16-octet group identifier, a 16-octet host UUID, a 1-octet service identifier, and a 2-octet port number in network byte order (big-endian):

{\small
\begin{verbatim}
  +------+------------+-----------+---------+------+
  | type | group UUID | host UUID | service | port |
  +------+------------+-----------+---------+------+
\end{verbatim}
}

The message type can either be a request~(\texttt{0x01}), an offer~(\texttt{0x02}), or a depart message~(\texttt{0x03}). Possible values for the service identifier are not specified in CHIRP itself but in the corresponding protocols for the services to avoid updating the protocol when a new service is introduced.

\subsection{Implementation}

CHIRP has been implemented independently in \mbox{C\texttt{++}} and Python~\cite{ConstellationGitLab}, as well as for the ESP32 microchip~\cite{MicroSatGitLab}. The protocol and its implementations have been tested in test beam environments.

Usually, only one program can bind a specific network port. To allow multiple satellites on the same machine to use CHIRP on port 7123, the \texttt{SO\_REUSEADDR} socket option for UDP was enabled.

One challenge faced during the implementation was machines with multiple active network interfaces. Using the default broadcast address (\texttt{255.255.255.255}) does not result in broadcasts sent to all network interfaces, which resulted in some services not being found. A solution for this is to iterate over all network interfaces and broadcast using the respective interface-specific broadcast addresses.

\begin{figure}
  \centering
  \includegraphics[width=0.95\linewidth]{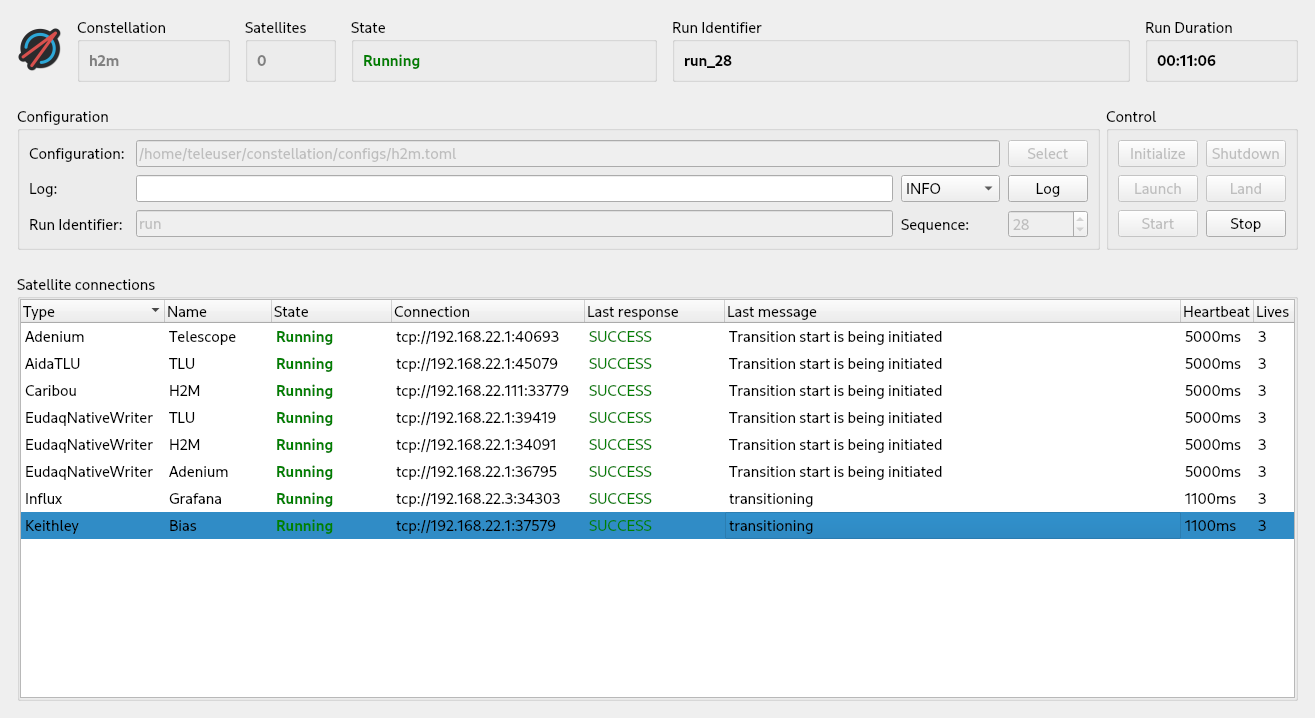}
  \caption{Screenshot of a user interface for Constellation}
  \label{fig:MissonControlH2M}
\end{figure}

In \autoref{fig:MissonControlH2M} a screenshot of a user interface for Constellation is shown. The screenshot was taken during a test beam where eight satellites were used in total. The user interface used CHIRP to discover the satellites that were running on three different machines across the local network. Neither the satellites nor the user interface required any configuration of IP addresses.

\section{Summary \& Outlook}

A network discovery protocol has been developed for a network-distributed control and data acquisition framework named Constellation. The protocol allows to discover constituents of the network without requiring prior knowledge of their IP addresses using UDP broadcasts as underlying network technology. The protocol includes a group identifier which allows to carry out multiple separate setups in the same local network. It also foresees dynamic allocation of network ports to allow running multiple instances on the same machine. Discovery can be achieved both as an early joiner and a late joiner to a network through a request and offer pattern.

Further improvements to the protocol are planned after extensive use in practice. Replacing UDP broadcasts with multicasts allows for use in networks where broadcasting is restricted from the router and reduces network traffic to non-participating machines. Using arbitrary length names instead of fixed length UUIDs improves the readability of log messages regarding the protocol. Lastly, leveraging a standardized serialization format like MessagePack~\cite{MessagePack} instead of defining bytes directly will result in simplified code paths.

\bibliographystyle{elsarticle-num} 
\bibliography{references.bib}

\end{document}